\documentclass[preprint, 12pt]{elsarticle}
\usepackage{hyperref}
\usepackage{lineno}
\usepackage{amsmath}
\usepackage{subfigure}
\usepackage{here}

\usepackage{docmute}

\newcommand*\patchAmsMathEnvironmentForLineno[1]{
  \expandafter\let\csname old#1\expandafter\endcsname\csname #1\endcsname
  \expandafter\let\csname oldend#1\expandafter\endcsname\csname end#1\endcsname
  \renewenvironment{#1}
     {\linenomath\csname old#1\endcsname}
     {\csname oldend#1\endcsname\endlinenomath}}
\newcommand*\patchBothAmsMathEnvironmentsForLineno[1]{
  \patchAmsMathEnvironmentForLineno{#1}
  \patchAmsMathEnvironmentForLineno{#1*}}
\AtBeginDocument{
\patchBothAmsMathEnvironmentsForLineno{equation}
\patchBothAmsMathEnvironmentsForLineno{align}
\patchBothAmsMathEnvironmentsForLineno{flalign}
\patchBothAmsMathEnvironmentsForLineno{alignat}
\patchBothAmsMathEnvironmentsForLineno{gather}
\patchBothAmsMathEnvironmentsForLineno{multline}
}

\usepackage{setspace}

\begin{document}

\begin{frontmatter}
\title{Evaluation of the X-ray SOI pixel detector with the on-chip ADC}

\author[ut]{Hiroumi~Matsuhashi\corref{cor1}}
\author[ut]{Kouichi~Hagino}
\author[ut]{Aya~Bamba}
\author[miyazaki]{Ayaki~Takeda}
\author[miyazaki]{Masataka~Yukumoto}
\author[miyazaki]{Koji~Mori}
\author[miyazaki]{Yusuke~Nishioka}
\author[kyoto]{Takeshi~Go~Tsuru}
\author[kyoto]{Mizuki~Uenomachi}
\author[kyoto]{Tomonori~Ikeda}
\author[kyoto]{Masamune~Matsuda}
\author[kyoto]{Takuto~Narita}
\author[jaxa]{Hiromasa~Suzuki}
\author[konan]{Takaaki~Tanaka}
\author[DS]{Ikuo~Kurachi}
\author[tus]{Takayoshi~Kohmura}
\author[tus]{Yusuke~Uchida}
\author[kek]{Yasuo~Arai}
\author[sizuoka]{Shoji~Kawahito}
\cortext[cor1]{Email Address: hiroumi.matsuhashi@phys.s.u-tokyo.ac.jp}
\address[ut]{Department of Physics, University of Tokyo, 7-3-1 Hongo, Bunkyo, Tokyo 113-0033, Japan}
\address[miyazaki]{Department of Applied Physics, Faculty of Engineering, University of Miyazaki, 1-1 Gakuen-Kibanadai-Nishi, Miyazaki, 889-2192, Japan}
\address[kyoto]{Department of Physics, Graduate School of Science, Kyoto University, Kitashirakawa Oiwake-cho, Sakyo-ku, Kyoto 606-8502, Japan}
\address[jaxa]{Institute of Space and Astronautical Science, Japan Aerospace Exploration Agency, 3-1-1
Yoshinodai, Chuo-ku, Sagamihara, Kanagawa 252-5210, Japan}
\address[konan]{Department of Physics, Faculty of Science and Engineering, Konan University, 8-9-1 Okamoto, Higashinada, Kobe, Hyogo 658-8501, Japan}
\address[DS]{D$\rm{\&}$S Inc., 774-3-213 Higashiasakawacho, Hachioji, Tokyo 193-0834, Japan}
\address[tus]{Department of Physics, Faculty of Science and Technology, Tokyo University of Science, 2641 Yamazaki, Noda, Chiba 278-8510, Japan}
\address[kek]{Institute of Particle and Nuclear Studies, High Energy Accelerator Research Org., KEK, 1-1 Oho, Tsukuba 305-0801, Japan}
\address[sizuoka]{Research Institute of Electronics, Shizuoka University, Johoku 3-5-1, Naka-ku, Hamamatsu, Shizuoka 432-8011, Japan}
\begin{abstract}

XRPIX is the monolithic X-ray SOI (silicon-on-insulator) pixel detector, which has a time resolution better than 10 $\rm{\mu}$s as well as a high detection efficiency for X-rays above 10 keV.
XRPIX is planned to be installed on future X-ray satellites.
To mount on satellites, it is essential that the ADC (analog-to-digital converter) be implemented on the detector because such peripheral circuits must be as compact as possible to achieve a large imaging area in the limited space in satellites.
Thus, we developed a new XRPIX device with the on-chip ADC, and evaluated its performances.
As the results, the integral non-linearity was evaluated to be 6 LSB (least significant bit), equivalent to 36~eV. The differential non-linearity was less than 0.7 LSB, and input noise from the on-chip ADC was 5~$\rm{e^{-}}$.
Also, we evaluated end-to-end performance including the sensor part as well as the on-chip ADC. As the results, energy resolution at 5.9~keV was 294 $\rm{\pm}$ 4~eV in full-width at half maximum for the best pixel. 

\end{abstract}

\end{frontmatter}

\section{Introduction}

Charge coupled devices (CCDs) have been widely used for X-ray detectors to observe celestial objects.
It is because of their good energy resolutions and imaging capabilities. For example, CCDs onboard the Hitomi satellite had an energy resolution of 160~eV at 6~keV \cite{CCDref}.
However, hard X-rays above 10~keV cannot be observed with CCDs due to high background above 6~keV \cite{suzaku}.
In neutron stars and black holes, particles accelerated to relativistic speeds emit X-ray radiation above 10~keV. As an example, in accreting neutron star, we are able to observe electron cyclotron resonance absorption lines above 10~keV. 
These have actually been observed \cite{cyclotron_sca}. The magnetic field of neutron star can be measured from this absorption line.  
Therefore, hard X-rays above 10~keV are an important probe for high-energy phenomena in the universe. 
Given this consideration, the detector should have low background and energy resolution above 10~keV.

\begin{figure}[tbp]
\centering
\includegraphics[scale=0.2]{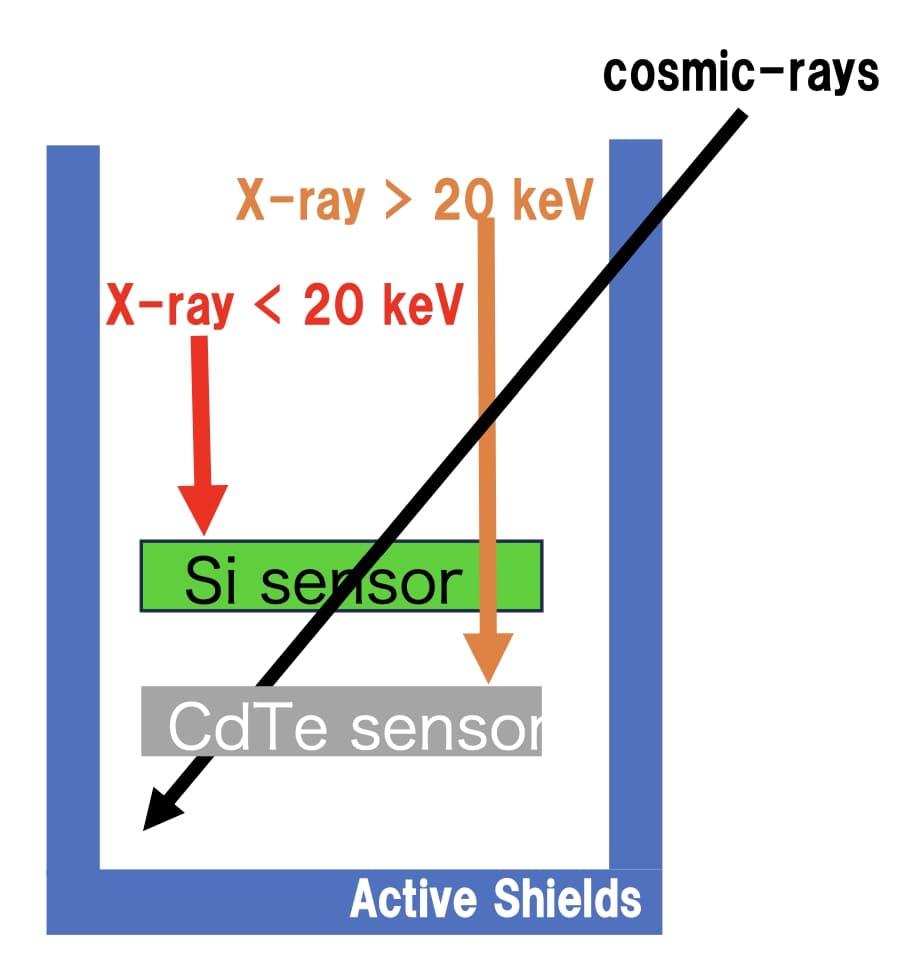}
\caption{A concept of hard X-ray detector system. To adopt the anti-coincidence method, two detectors are surrounded by active shields.}
\label{fig1}
\end{figure}

To realize the hard X-ray observation, reduction of background above 10~keV is one of the most important issues.
To cope with background, we adopt a detector structure as shown in Fig.~\ref{fig1}.
One of the two detectors is a silicon sensor detecting X-rays below 20~keV, and the other one is a cadmium telluride sensor detecting above 20~keV.
In this structure, two detectors are surrounded by an active shield to employ the anti-coincidence method \cite{FORCE,FORCE2}. 
In the anti-coincidence method, when the detector and active shield detect an event simultaneously, the event is discarded as a background event.
This method efficiently reduces the background above 10 keV because it is dominated by cosmic-ray particles, which usually hits both the detector and shield.
The active shield is known to trigger at 10 kHz \cite{NXBtime}, hence the detector should have high time resolution to adopt the anti-coincidence method, which is not possible with CCD because its time resolution is a few seconds. The detector requires high time resolution in order to adopt this method. Then, we can reduce background above 10~keV and observe wide-band X-rays.

We have been developing XRPIX as a future X-ray detector to realize the wide-band X-ray observation \cite{tsuru2018}. 
Our goal specification is shown in table \ref{tabSPEC}.
XRPIX is the monolithic X-ray detector in which the readout circuit and the silicon sensor layer are integrated with an insulation layer of $\rm{SiO_{2}}$ in between by utilizing the silicon-on-insulator (SOI) technology.
We are developing XRPIX to achieve the specifications presented in Table \ref{tabSPEC}.
By implementing the self-trigger function in pixel circuits, the time resolution of XRPIX is better than 10~$\rm{\mu}$s \cite{tsuru2018,timeRES10us}. This high time resolution allows use of anti-coincidence method. 
This method enables XRPIX to reduce the background at 20~keV to 1/100 of CCD \cite{tsuru2018}.
Also, since the sensor layer is made of high resistivity silicon using the SOI technology, XRPIX’s sensor layer is thicker than 300~$\rm{\mu}$m. It can catch X-rays up to 20 keV efficiently.  
Goal of energy resolution is 140 eV at 6 keV which is comparable to CCD performance \cite{tsuru2018}. By this performance, we can resolve X-ray spectrum precisely. 
Thus, XRPIX is the most promising detector for the future wide-band X-ray observation.

In previous studies \cite{harada,takeda}, XRPIX has succeeded in achieving an energy resolution better than 300~eV at 6~keV. 
For soft X-rays below 1 keV, XRPIX is able to detect X-rays at 0.68 keV from previous studies \cite{soft_Xray}.
 At present, the current XRPIX is inferior to CCDs in terms of energy resolution, but a better resolution of 140 eV was achieved in the previous XRPIX \cite{xrpix6e}. Since this is thought to be due to the difference in wafers, the resolution of 140 eV is quite achievable.

In addition to the performance improvements described above, it is equally important to maintain the size of a sensor board as small as possible, in which XRPIX and external peripheral circuits co-exist. 
This is especially true in our detector design in Fig.~\ref{fig1}; a larger sensor board requires a bigger and heavier shield, which could be a significant disadvantage because the space and mass allocated to a detector are in general quite limited in a satellite. 
For example, the sensor board of the HXI onboard Hitomi, which adopted the same design of Fig.~\ref{fig1}, was limited to a size of a 10 cm square \cite{hitomiex}.  
One effective way to make the imaging area of XRPIX as large as possible and also keep the size of a sensor board as adopted in the HXI is to implement some functionalities of the external peripheral circuits into XRPIX.

Therefore, we developed a new XRPIX device,  in which analog-to-digital converters (ADCs) are installed. 
In this paper, we report the specifications and performance of XRPIX with the on-chip ADC. In section~2, we describe XRPIX and ADC specifications. In section~3 and 4, we present the experimental setup and results of the performance evaluations of XRPIX. 
In section~5, we discuss basic performance of ADCs from a perspective of astronomical observation. In section~6, we summarized performances of the on-chip ADC and XRPIX.

\begin{table*}[tbp]
    \centering
    \begin{center}
    \caption{Specifications of XRPIX}
    \begin{tabular}{lccl}
        \hline
        \hline
        \: & requirement & goal & current status \\ \hline
        Time resolution & 10 $\rm{\mu}$s  &10 $\rm{\mu}$s & achievable by design. \\
        Lowest detectable energy & 1.0 keV & 0.3 keV & 0.68 keV \cite{soft_Xray} \\
        Sensor layer & 300 $\rm{\mu}$m & 500 $\rm{\mu}$m & 300 $\rm{\mu}$m  \\
        Energy resolution at 6~keV& 300 eV & 140 eV & 140 eV \cite{xrpix6e} \\
        \hline
    \end{tabular}
    \end{center}
    \label{tabSPEC}
\end{table*}

\section{Designs of XRPIX9 with the on-chip ADC}

\begin{figure}[tbp]
\centering
\includegraphics[scale=0.35]{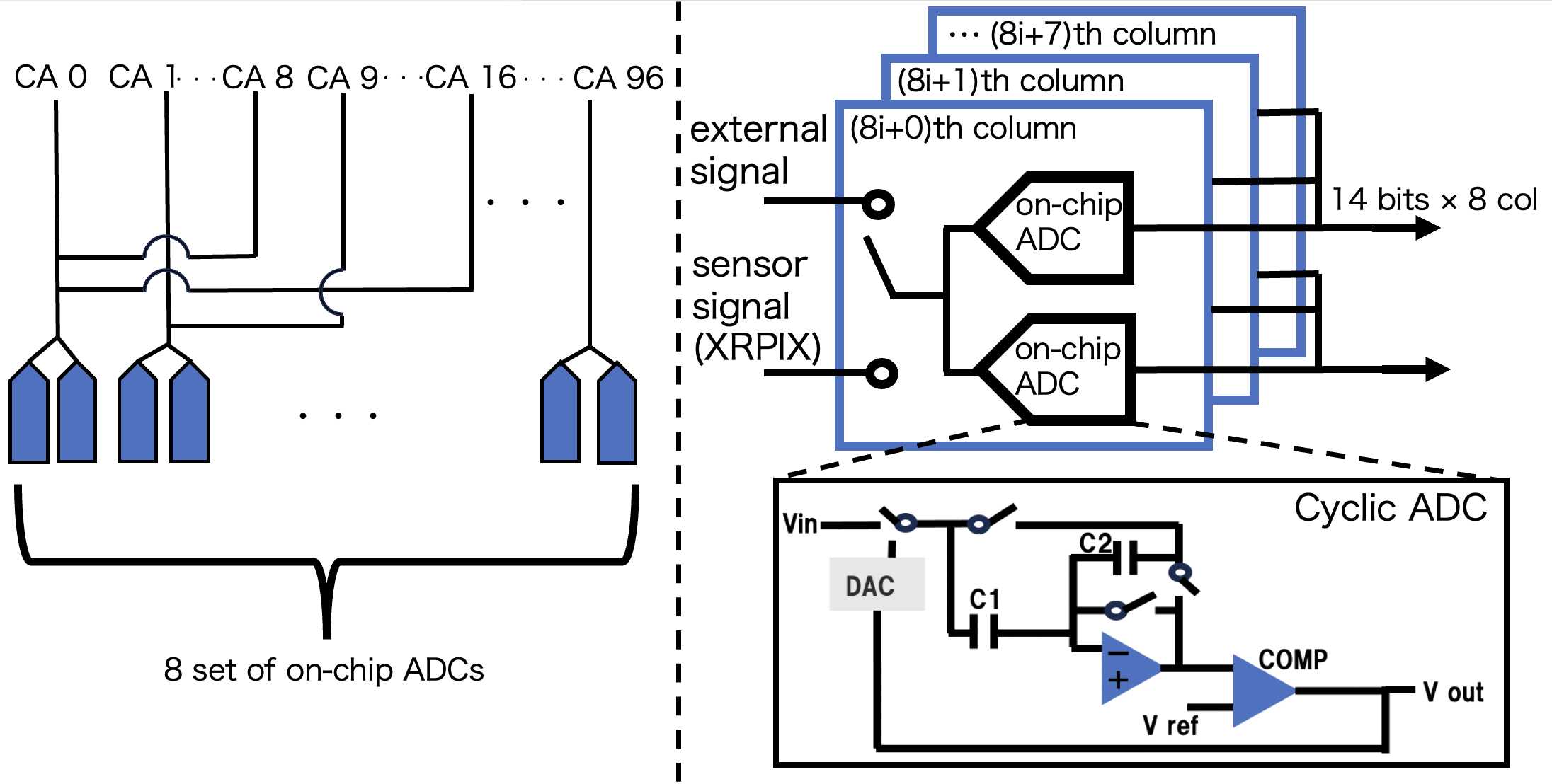}
\caption{Schematic of the ADC on XRPIX9. Left panel : Each column is connected to the on-chip ADC. Each ADC is connected in 8 column cycles. Right panel : The ADC can be connected to external signals (e.g. function generators) and sensor signals. One column has two ADCs.}
\label{fig2}
\end{figure}

Table~\ref{tabADC} provides an overview of the XRPIX9 sensor. 
The structure of the sensor part of XRPIX9 is the same as that of XRPIX8 \cite{yukumoto}. 
The 300~$\rm{\mu}$m thick sensor layer enables observations up to 20~keV. The pixel size is as small as 36~$\rm{\mu}$m, and the number of pixels is 64$\rm{\times}$96. 
These pixels are divided into 6 test element groups (TEG) of $\rm{32\:\times32}$ pixels with slightly different doping conditions and circuit configuration.
Fig.~\ref{fig2} shows the schematic of the ADC on XRPIX.
In Fig.~\ref{fig2}, index $i$ runs from 0 to 11, covering all 96~columns, for $i$ is 0, column address (CA) $=$~0,~1,~...,~7, are read by ADC0,~1,~...,~7, respectively.
For $i\:=$ 1, CA $=$ 8,~9,~...,~15 will be read by ADC0,~1,~...,~7, respectively.
Therefore, CA $=$~0,~1,~...,~7, and 8,~9,~...,~15, are read by the same ADCs and 96 columns are connected in 8~cycles to 16 ADCs with 2 ADCs per column as shown in Fig.~\ref{fig2}.
One column has two ADCs for circuit redundancy, and we used only one ADC for each column in the evaluation of X-ray performance described in following sections.

The schematic structure of the on-chip ADC is shown in the inset panel at the bottom of Fig.~\ref{fig2}. 
The type of ADC is the cyclic ADC.
The cyclic ADC consists of a switched capacitor amplifier and a 1.5-bit comparator \cite{kawahito,Park}.
AD conversion is performed by repeating amplification and comparation for the number of bits, so the conversion can be performed in a small area and at high speed. 
The cyclic ADC on XRPIX9 is able to convert analog to digital signals within 6~$\rm{\mu}$s. 
AD conversion of 0.4 --1.5 V at 14 bits allows XRPIX to cover a wide energy bandwidth of about 100 keV with a typical gain~=~40~$\rm{\mu}$V/$\rm{e^{-}}$.

\begin{table}[tbp]
    \centering
    \caption{Designs of XRPIX9 and the on-chip ADC}
    \begin{tabular}{ccr}
        \hline
        \hline
        Sensor layer & 300 $\rm{\mu}$m  \\
        Pixel size & 36 $\rm{\mu}$m $\times$ 36 $\rm{\mu}$m  \\
        Number of pixels &64$\times$96 \\
        \hline
        On-chip ADC type & Cyclic ADC  \\
        ADC size & 20 $\rm{\mu}$m $\times$ 2 mm  \\
        AD conversion time & 5.96 $\rm{\mu}$s  \\
        Number of bits & 14 bit  \\
        Number of ADC units & 16 units  \\
        Voltage range & $\rm{0.4\: - \: 1.5\: V}$ \\ 
        Resolution of voltage & 67 $\rm{\mu}$V \\ \hline
    \end{tabular}
    \label{tabADC}
\end{table}

\section{Unit test of the on-chip ADC}
Before evaluating the end-to-end performance of XRPIX9 from the X-ray sensor part to the on-chip ADC, we conducted a unit test of the on-chip ADC to evaluate its performance. 
To evaluate the performance of the ADC, we input external signals to ADCs from the function generator (model:{\it{RIGOL DG2102}}). 
The vertical resolution of this function generator is 16~bit. 
We inputted ramp wave and constant voltage to evaluate the integral non-linearity (INL), differential non-linearity (DNL), and noise. 

\begin{figure}[tbp]
\centering
\includegraphics[scale=0.43]{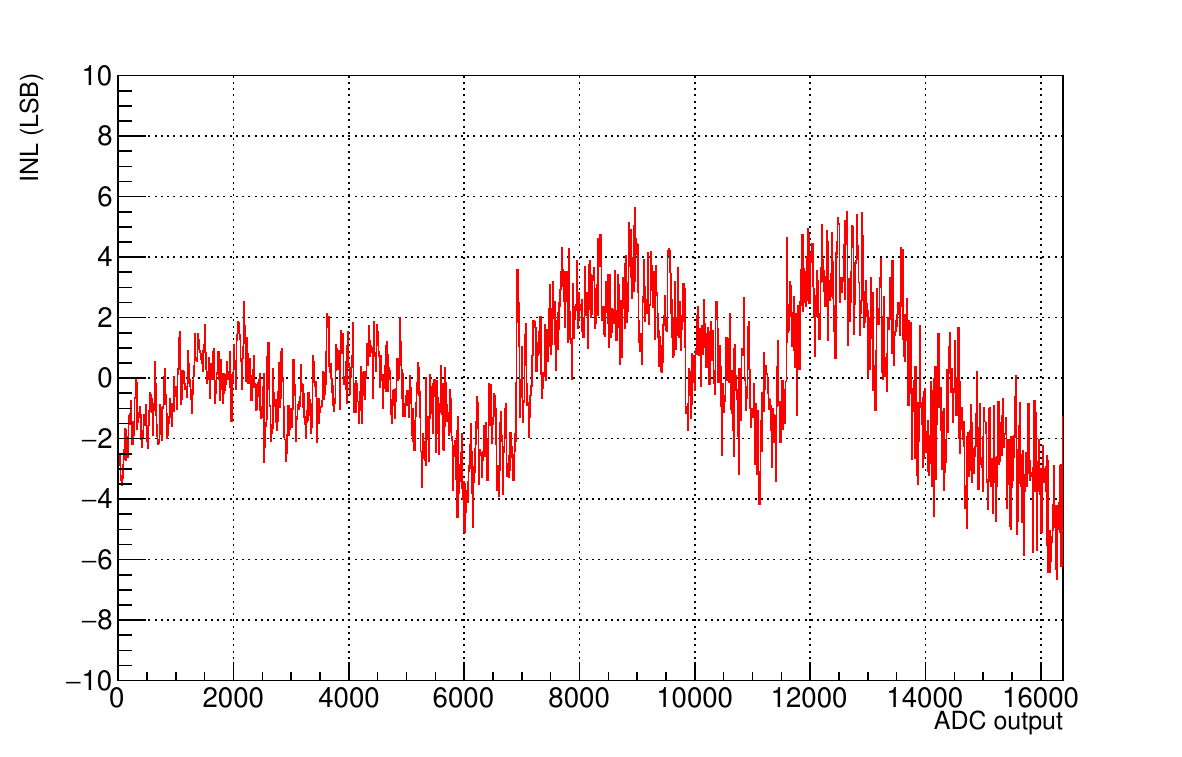}
\caption{Measured INL of the ADC of XRPIX9. Though this figure shows only one ADC, the other 15 ADCs have similar performance. }
\label{fig3}
\end{figure}

We evaluated the INL first. A 100~mHz ramp wave was input from the function generator to the ADC and the output waveform was fitted with a straight line. 
Residuals between the best-fit straight line and the output waveform correspond to the INL. 
Fig.~\ref{fig3} shows the measured INL of the on-chip ADC of XRPIX9. The INL is worse around 7,000, 12,000, and 16,000~channles, with maximum values about 6~LSB (least significant bit). 
As explained in detail in section 5.1, these jump shapes around 7,000, 12,000 and 16,000~channels are caused by cyclic ADC configuration. 
In X-ray energy, the INL of 6~LSB corresponds to 36~eV using the typical conversion gain of $\rm{40\:\mu V/e^{-} }$ in the latest XRPIX.
Thus, the INL has no critical adverse effect on energy resolution of 300~eV at 6~keV.

\begin{figure}[tbp]
\centering
\includegraphics[scale=0.43]{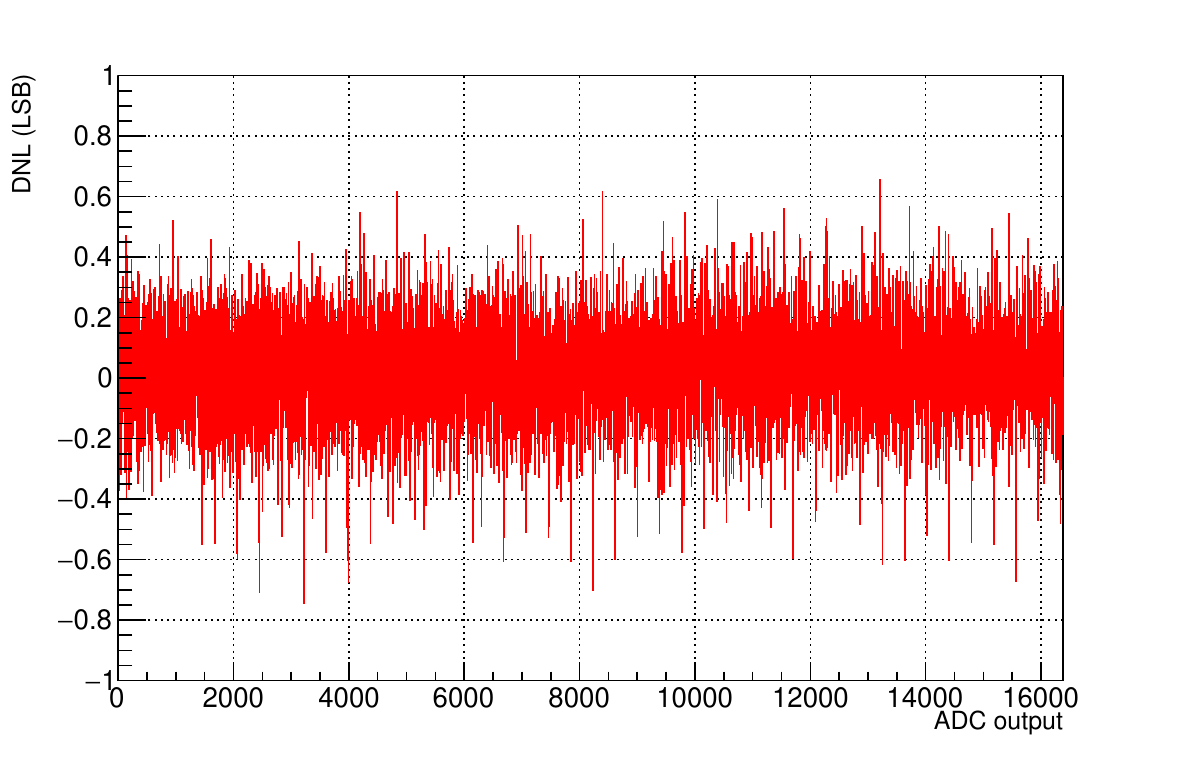}
\caption{Measured DNL of the ADC of XRPIX9. The DNL contains a Poisson fluctuation of $\rm{\pm0.05\:LSB}$. Though this figure shows only one ADC, the other 15 ADCs have similar performance.}
\label{fig4}
\end{figure}

Next, we evaluated the DNL by inputting a 2 mHz ramp wave to the ADC. 
The ramp wave provides a uniform input in voltage space. 
Hence, for an ideal ADC, where the step width is uniform in voltage space, a histogram of AD outputs is a flat histogram with perfectly constant values. 
On the other hand, for actual ADCs, the histogram is lower than the ideal flat histogram at channels with narrow step widths. 
Since the DNL refers to the non-uniformity of the step width in the voltage space, the DNL can be evaluated with such histograms. 
Therefore, we histogrammed AD outputs and evaluated deviations from the ideal flat histogram.
Fig.~\ref{fig4} shows the DNL of the on-chip ADC of XRPIX9, and it is less than 0.7~LSB.
It means that the ADC outputs all channels without missing any channels.

\begin{figure}[tbp]
\centering
\includegraphics[scale=0.4]{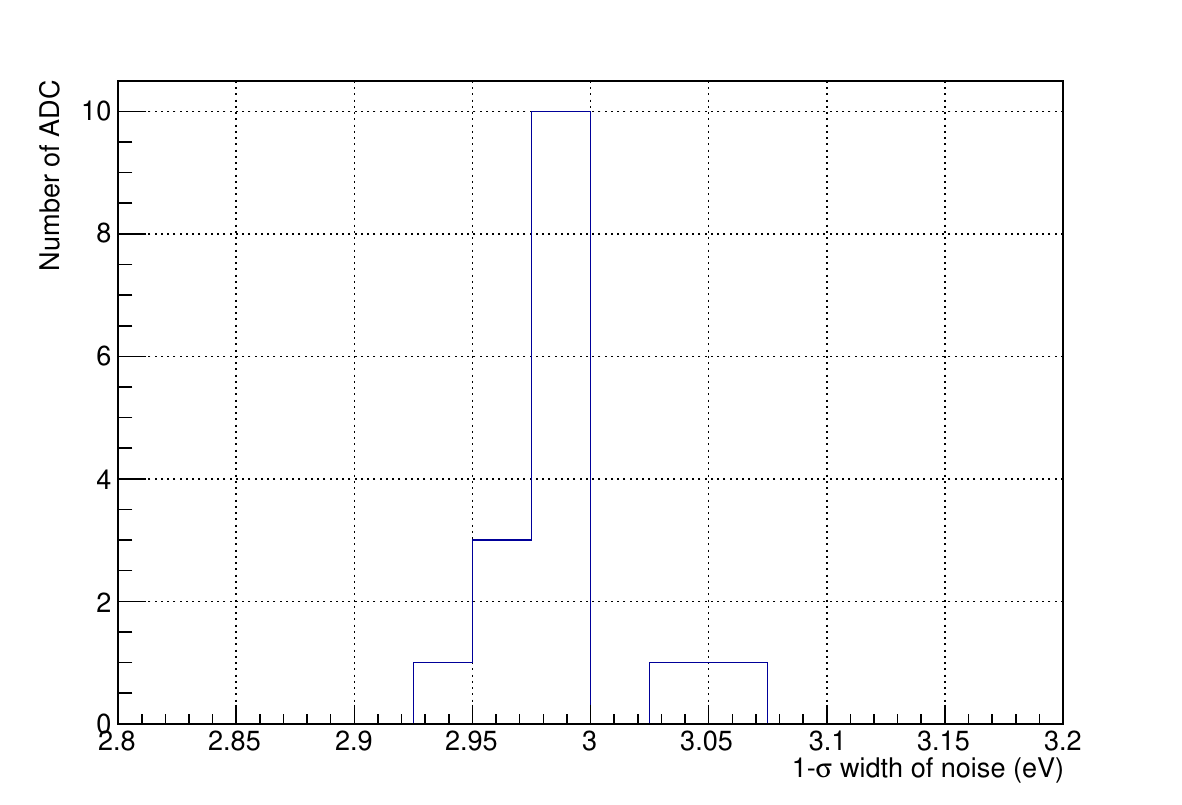}
\caption{Histogram of the noise (1 $\rm{-\:\sigma}$) in each ADC of XRPIX9..
}
\label{fig5}
\end{figure}

Finally, we evaluated the noise of the ADC by inputting constant voltage of 1.0~V to the ADCs. 
This input constant voltage contains the noise of $\rm{1.13\: \pm \: 0.02 \: e^{-}}$ from the function generator.
We evaluated the standard deviation of each of the 16 ADC outputs.
In Fig.~\ref{fig5}, we histogramed the standard deviation as the noise of each ADC.
Fig.~\ref{fig5} shows that the noise of most ADCs (1$\sigma$) is 3~LSB, corresponding to 5~$\rm{e^{-}}$.  As with the INL, it indicates that the ADC-derived noise is negligible level considering that the energy resolution is 300~eV.

\section{XRPIX9 end-to-end test}
We evaluated end-to-end performance. In this experiment, XRPIX9 was irradiated with X-rays from $\rm{^{55}Fe\:and\:^{109}Cd}$.
In this evaluation, a TEG with the same doping and circuit configurations as the default XRPIX8 are used for evaluation \cite{yukumoto}.
XRPIX9 was cooled down to $\rm{-60}$~°C in thermostatic chamber to reduce the shot noise of the dark current.
We applied a backbias voltage of $\rm{-180}$~V to XRPIX9 for full depletion.
The voltage of the buried p-well at the oxide interface was set to $\rm{-2.5}$ V as in the previous study \cite{yukumoto}. This p-well reduces the dark current from the interface. 
We irradiated X-rays from radioisotopes ($\rm{^{55}Fe}$ and $\rm{^{109}Cd}$) on XRPIX9 and X-ray events were obtained via the on-chip ADC.
The event data were acquired only in 8×8 pixels located in the TEG with default XRPIX8 condition to efficiently obtain the X-ray events.

\begin{figure}[tbp]
\centering
\includegraphics[scale=0.43]{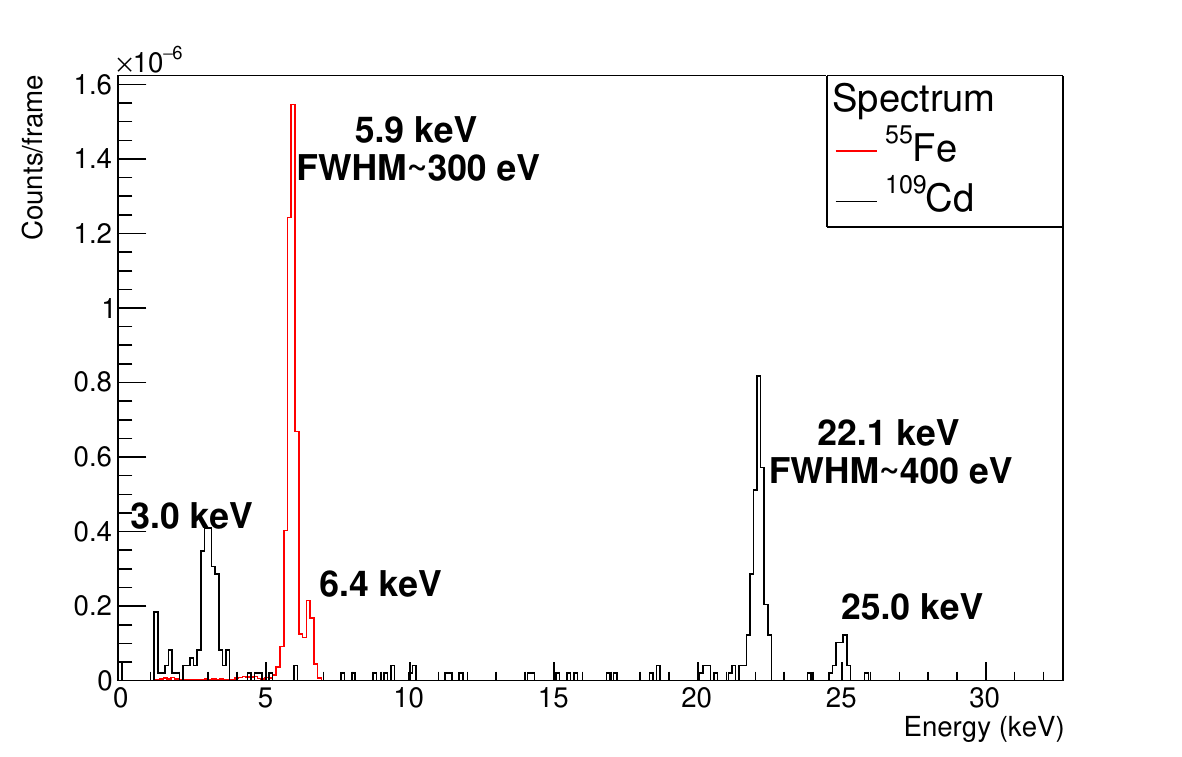}
\caption{X-ray spectrum of XRPIX9 read via the on-chip ADC. Red spectrum were obtained with $\rm{^{55}Fe}$ and black spectrum with $\rm{^{109}Cd}$. In this spectrum, only single events were extracted by $\rm{8\times8}$ pixels. }
\label{fig6}
\end{figure}

We succeeded in acquiring broadband X-ray spectra from XRPIX9 via the on-chip ADC as shown in Fig.~\ref{fig6}. 
To evaluate the spectral performance without calibrating the pixel-by-pixel gain variation, we used only single events, in which one pixel exceeds a threshold of 1050~eV and the surrounding 3×3 pixels do not exceed the other threshold of 220~eV.
As the results, the energy resolution was $\rm{296 \pm 4}$~eV at 5.9~keV and $\rm{366 \pm 49}$~eV at 22.1~keV, consistent with previous studies of 300~eV at 6~keV.

\section{Discussion}
\subsection{Basic Performance of ADCs from a perspective of Astronomical Observation}
In X-ray astronomical observations, local INL features have a considerable impact on the local spectral features such as emission and absorption lines.
In Fig.~\ref{fig3}, the INL plot has two jumps around 7000~channels and 12,000~channels.
This results in multiple AD outputs for a single energy. Thus, the energy resolution is worse around these specific channels. 
Though local structures can be reduced by coadding outputs from different ADCs in general, it is not the case for our cyclic ADC in XRPIX9. 
The large jumps around 7000 and 12,000~channels are inevitably caused by the 1.5-bit stage configuration adopted in our cyclic ADC \cite{15bitADC}.
Alternatively, it can be handled by changing the offset level of signal to move the jumps to energy bands where there are no emission lines from celestial objects.

\begin{figure}[tbp]
\centering
\includegraphics[scale=0.4]{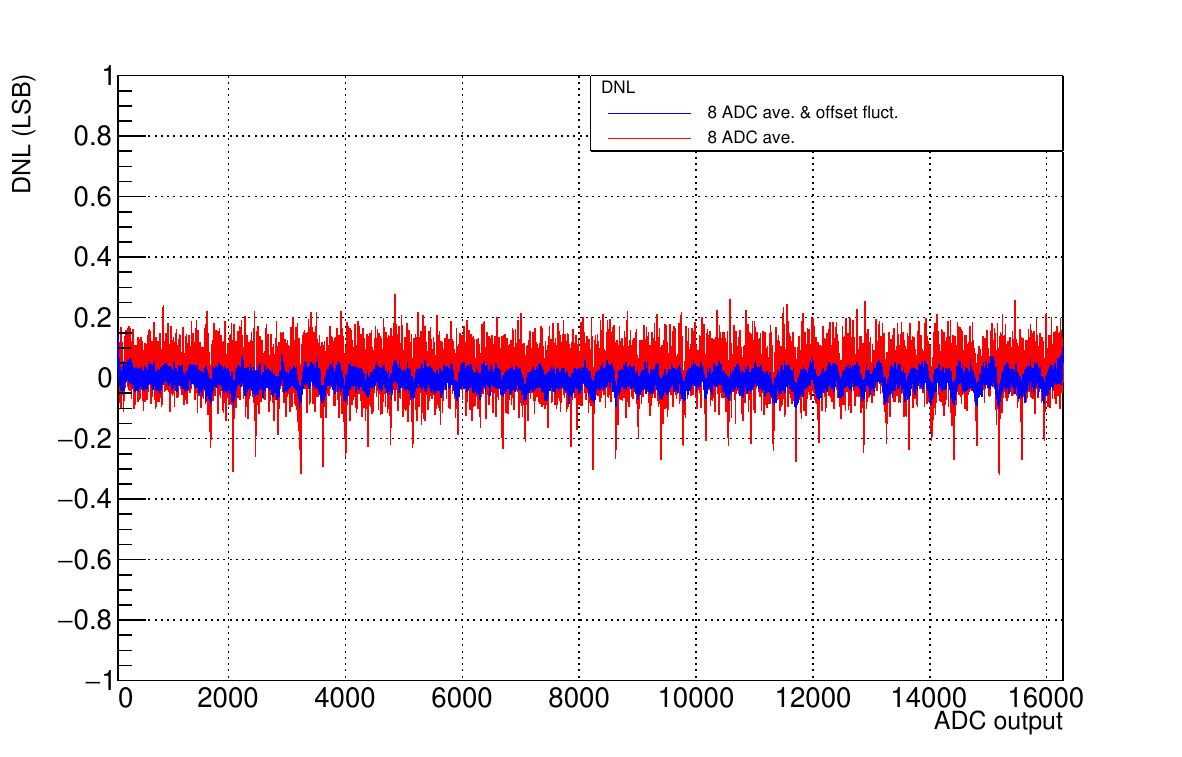}
\caption{The improved DNL under the realistic observational situation. 
We evaluated the DNL averaged by 8 ADCs with signal offset fluctuations for each event.}
\label{gaus}
\end{figure}

The DNL directly affects spectroscopic observations of the continuous component of X-ray spectrum, commonly observed in the hard X-ray band. 
Fig.~\ref{fig4} is equivalent to observing continuous X-ray emission in an astronomical observation.
Fig.~\ref{fig4} clearly shows local negative DNLs of up to $\rm{-0.7}$~LSB. 
It means that there will be artificial spectral drops at specific channels on the X-ray spectrum.
However, under the more realistic situation, the effect of DNL would be significantly suppressed. 
By considering an event-by-event fluctuation of the offset level and assuming to extract spectra from more than eight columns, the DNL decreases to $\rm{\pm 0.1}$~LSB, as shown in Fig.~\ref{gaus}.
These results show that the performance of the DNL is not a problem for astronomical observations.

\subsection{Comparison between an external ADC and the on-chip ADC}
\begin{figure}[tbp]
\centering
\includegraphics[scale=0.4]{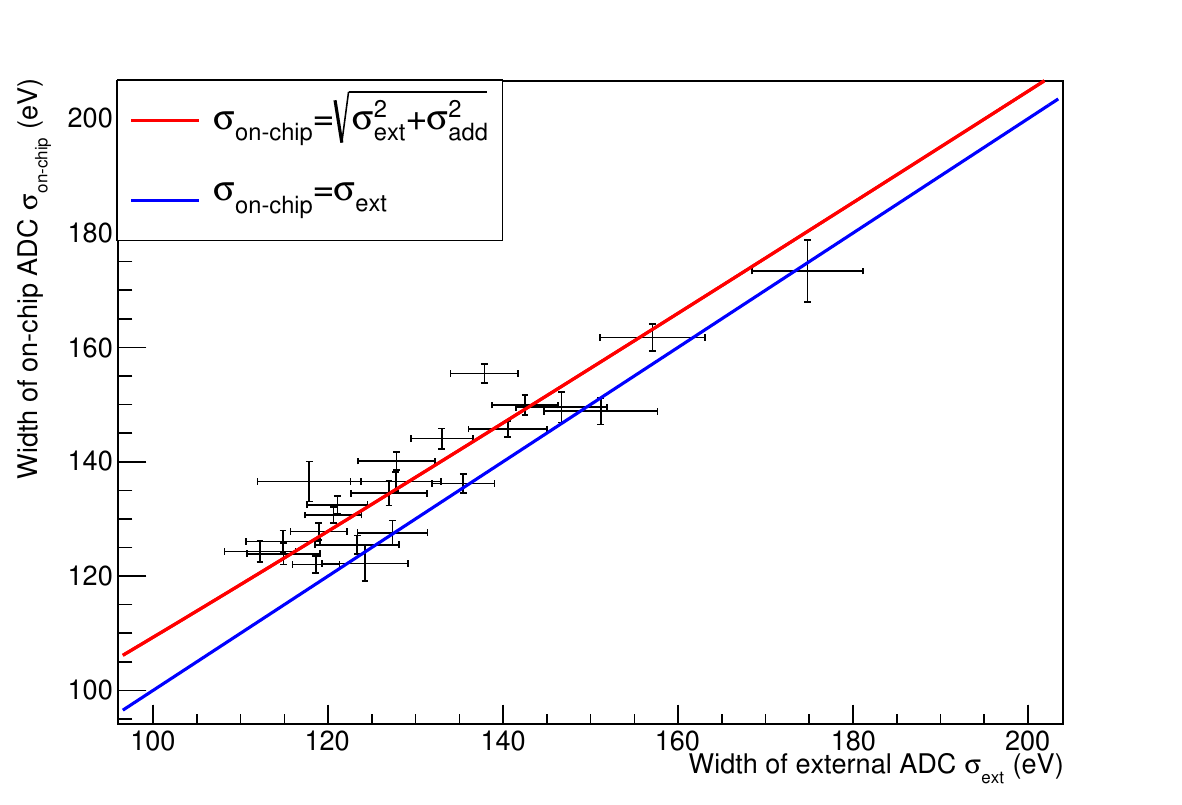}
\caption{A correlation between energy resolutions of the external ADC $\sigma_{\textrm{ext}}$ and that of the on-chip ADC $\sigma_{\textrm{on-chip}}$. 
As the energy resolution, 1-$\rm{\sigma}$ width at 5.9 keV was adopted. 
}
\label{fig10}
\end{figure}

We more quantitatively evaluated the spectral performance by comparing on-chip and external ADCs.
 All the previous studies of the XRPIX series used an external ADC. An external ADC is installed on the readout board. 
We compared the width in the standard deviation of 5.9~keV line from $\rm{^{55}Fe}$. 
We defined the width of an on-chip ADC and an external ADC as $\rm{\sigma_{\textrm{on-chip}}}$ and $\rm{\sigma_{\textrm{ext}}}$, respectively.
Fig.~\ref{fig10} shows a correlation between $\rm{\sigma_{\textrm{on-chip}}}$ and $\rm{\sigma_{ext}}$ in each pixel.
From Fig. \ref{fig10}, most of the pixels are distributed along $\rm{\sigma_{\textrm{on-chip}}=\sigma_{ext}}$ (blue line). This result show that there is no crucial performance difference between an external ADC and an on-chip ADC.

A more careful investigation indicates that the data points locate slightly above the blue line.
We estimated an additional noise originating from the on-chip ADC by convolving it with $\rm{\sigma_{ext}}$ as 

\begin{equation}
\sigma_{\textrm{on-chip}} = \sqrt{\sigma_\textrm{ext}^{2}+\sigma_\textrm{add}^{2}}.
\label{fitting}
\end{equation}
Here, we assumed that all noise follows gaussian distributions.
By fitting with Eq. 1 (red line in Fig.~\ref{fig10}),  the additional noise $\rm{\sigma_{add}}$ was found to be $\rm{44 \pm 3}$~eV.  
To investigate whether $\rm{\sigma_{add}}$ is energy dependent, a similar measurement was made at 22.1 keV and found to be $\rm{\sigma_{add}}$ = 58 $\rm{\pm}$ 7~eV.
Thus, there is no clear energy dependence.
 The noise of ADC of $\rm{5\:e^{-}=18\:eV}$ described in Sec.~3 cannot explain this additional noise. 
The cause is still unknown, but one possible cause is interference between the on-chip ADC and other circuits. This additional noise will be investigated more in future.

\begin{figure}[tbp]
\centering
\includegraphics[scale=0.4]{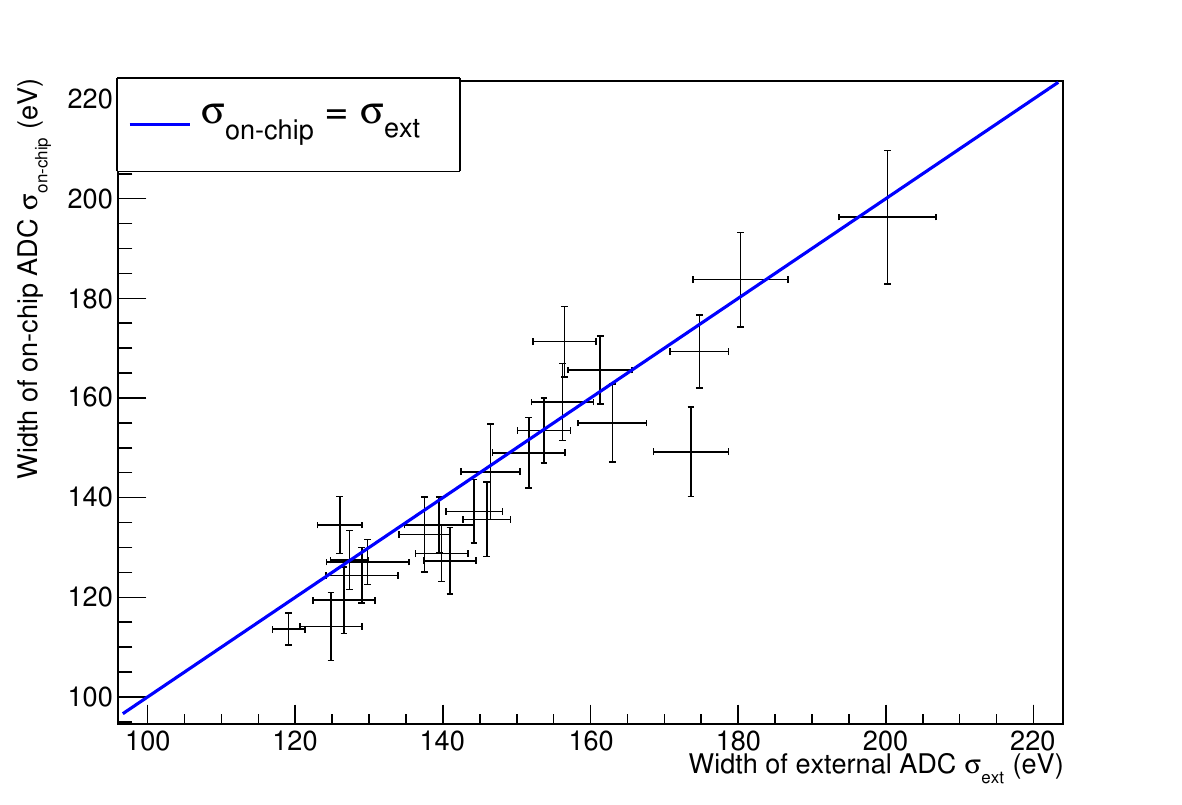}
\caption{
Same as Fig. 8, but $\sigma_{\textrm{on-chip}}$ was measured by increasing the gain of the pixel output signals by a factor of 4.
}
\label{fig noise}
\end{figure}

This additional noise is practically compensated by increasing the gain of the pixel output signal. We re-evaluated the energy resolution with the on-chip ADC by increasing the gain by a factor of 4. As a result, the additional noise was reduced as shown in Fig.~\ref{fig noise}. Quantitatively, we obtained an upper limit of the additional noise of 10~eV from the fitting with Eq.~1. 
 Here, changing the gain of the pixel signal does not change INL and DNL. However, in the X-ray energy scale, INL decreases by 1/4 when the pixel signal increases by a factor of 4.

\section{Coclusion}
We developed and evaluated XRPIX9, which is the first device in XRPIX series equipped with the on-chip cyclic ADC. 
We evaluated the performance of  the ADC unit by inputting external signals. We found that the INL, DNL, and noise were 6~LSB, $\rm{- 0.7}$~LSB, and $\rm{5\:e^{-}}$, respectively.
All of these have no critical adverse effect on X-ray astronomical observations. 
Also, we evaluated XRPIX9 including X-ray sensor part as well as the on-chip ADC.  We successfully obtained spectrum from the on-chip ADC.
Then, we found that the energy resolution was as good as $\rm{294 \pm 4}$~eV at 5.9~keV.
We compared the width (1$\rm{\sigma}$) of 5.9~keV emission line from $\rm{^{55}Fe}$ between the on-chip ADC and an external ADC, which was used all the previous studies of XRPIX.
As the result, there was $\rm{44 \pm 3}$~eV of noise from the on-chip ADC and noise due to the interference of the on-chip ADC.
By increasing the gain of the pixel output signals by a factor of 4 before inputting the ADC, the noise was suppressed to less than 10~eV.

\section{Acknowledgement}
The authors would like to thank XRPIX team for technical assistance with the experiments. 
We thank T. Kosugi and T. Iida of TOPPAN Holdings Inc. for designing the ADC and advising us on its evaluation.
This work was supported through the activities of VDEC, The University of Tokyo, in collaboration with Cadence Design Systems and NIHON SYNOPSYS G.K.
This work was supported by JSPS KAKENHI Grant Numbers JP22H01269, JP21K13963, JP23H04006, JP22H04572, JP21H05461, JP21H04493, JP21K18151, JP21H01095. 
We greatly thank and acknowledge the referees for their comments which helped us to improve and clarify the manuscript.

\bibliography{FORCE,harada,takeda,yukumoto,kawahito,15bitADC,hitomiex,NXBtime,FORCE2,tsuru2018,suzaku,CCDref,Park,cyclotron_sca,soft_Xray,xrpix6e,timeRES10us}


\bibliographystyle{elsarticle-num-names}

\end{document}